# The Wonderful Toy of 20th Century can be a Disaster in 21st Century:
## Scenario and Policies Regarding Mobile Waste in India

Neeta Sharma[1] and Manoj Kumar[2]

[1]Department of Computer Science, International Management Centre, New Delhi, India
[2]Department of Computer Science, Maharaja Surajmal Institute, New Delhi, India

*Abstract-* **The subscribers' base of mobile phones is increasing globally with a rapid rate. The sale of mobile phones has exceeded those of personal computers. India is the second largest telecommunication network in the world in terms of number of wireless connections after China. Telecom companies are ready to tap a large unexplored market in India with lucrative offerings. Smart phones sale are at its peak. 3G technology is also ready to play a lead role in mobile revolution. Due to the low average life of the mobile phones, lack of awareness among users and in absence of government policies, mobile waste is accumulating in vast amount in India. Without a proper system of recycling, the unsafe disposal is causing a variety of environmental and health problems.**

**This paper discusses the various issues related to the worldwide growth of mobile phones, the insecure methods of disposal and the regulations and policies in India. We intend to put forward some challenges and advices.**

## I. INTRODUCTION

Mobile phones are part of day-to-day life, keeping us in touch with the people we love or helping us keep up with events at the office. Every teenager, working person, and parent seems to have a mobile phone within arm's reach, and for good reason. They are not only convenient for keeping in contact, but also they have proven to be essential in emergencies. [7]

The first mobile phone was invented in 1973 by Dr. Martin Cooper at Motorola. The first commercially automated cellular network First Generation (1G) was launched in Japan by NTT (Nippon Telegraph & Telephone) in 1979. Several other countries also launched 1G network in the early 1980s including UK, Mexico and Canada. The technology in these early networks was pushed to the limit to accommodate increasing usage. [8]

In 1990s the Second Generation (2G) mobile phone systems emerged, primarily using the GSM (Global system for mobile communications) standard. 2G network is based on digital transmission rather than the analog transmission of 1G. 2G provides fast and out-of-band phone-to-network signaling. The first full internet service on mobile phones was introduced by NTT DoCoMo in Japan in 1999. Later, Industry began to work on Third Generation (3G) technology providing high speed IP data networks and mobile broadband. As the penetration of 2G and 3G phones have increased many folds in recent years, users are utilizing mobile phones in their daily lives. Trends show that there would be an ever increasing demand for greater data speeds. [8]

3GPP (3rd Generation Partnership Project) Long Term Evolution (LTE) is the latest standard in the mobile network technology tree that produced the GSM/EDGE (Enhanced Data rates for GSM Evolution) and UMTS/HSPA ( Universal Mobile Telecommunications System /Enhanced Data rates for GSM Evolution)network technologies. [22] Although LTE is often marketed as 4G which was first proposed by NTT DoCoMo of Japan and has been adopted as the international standard. [4]

The mobile industry has expanded its reach to every corner of the earth in recent past. Almost 90% of the entire earth is under the mobile coverage now.

The telecommunication sector continued to register significant success during past few years and has emerged as one of the key sectors responsible for India's economic growth. Today, India is the fastest growing telecom market in the world. Its population is growing, so is the number of mobile subscribers. India is flooded with telecom operators, wide range of handsets, lucrative offerings and low budget plans.

The unexpected growth of the mobile market may generate tones of mobile waste in near future. The awareness among the mobile users and mobile industry and the regulation and policies are the need of the hour.

## II. ENORMOUS GROWTH OF MOBILE PHONES

The usage of mobile phones has become ubiquitous in our daily lives. In developed regions every individual has a mobile phone and its penetration is drastically increasing in developing countries [21]. "People here seem to go out of the house with only their mobile phones and car keys". Today, it has become a Swiss Knife for though it might not have a corkscrew or a nail cutter but a mobile is a phone, a data handler and a one-stop gadget for all entertainment and communication needs. [20] According to the most recent data from the UN agency International Telecommunication Union (ITU), more than half the homes in regions such as Asia, South America than in Europe and North America, even in rural areas, have a mobile phone connection. [22]

### A. World's Scenario

The number of mobile phone subscribers has doubled in the past five years and is expected to grow by 10 percent to 5.6 billion in 2011. The growth in developing and emerging countries is especially strong. The threshold of 5 billion mobile phone subscribers will be exceeded this year for the first time. 800 million persons are already using the fast UMTS mobile communications standard; an increase of 37 per cent. In 2011, there will already be more than one billion UMTS subscribers. [21]

In EU, the number of mobile phone subscribers was expected to rise to around 650 million by the end of 2010 but it came around 906 million. Germany has the maximum number of mobile phone connections in the EU: around 111 million by the end of 2010. Germany is followed by Italy (87 million), Great Britain (81 million), France (62 million) and Spain (57 million). There are 220 million mobile subscribers in Russia and 287 million in the USA. The number is increasing more in Asia and South America than in Europe and North America. In China, the number of mobile phone subscribers has risen by almost 13 percent this year to around 844 million. This figure is expected to grow by one-tenth within the next year to 930 million. [22] Worldwide mobile device sales to end users totalled 1.6 billion units in 2010, a 31.8 percent increase from 2009 (Table I). According to Gartner Inc., smart phone sales to end users were up 72.1 percent from 2009 and accounted for 19 percent of total mobile communications device sales in 2010. [23]

*B. Indian Scenario*

India is the second largest telecommunication network in the world in terms of number of wireless connections after China with more than 752 million mobile phone subscribers in December, 2010. [24]

In recent times, mobile phones have gained remarkable popularity in consumer markets across India. India today serves as a lucrative market for all mobile phone manufacturers across the world. Apart from the big players like Nokia, Apple, RIM, HTC, Samsung, LG, Motorola and Sony Ericsson, the Indian mobile handset makers Lava, MicroMax, Spice, Karbonn, Videocon and Intex have flooded the Indian mobile market with wide variety of mobile handsets.

The popularity of smart phones is also growing and everybody seems to be interested to replace his old handset with fully featured smart phone. According to IDC India, the smart phones sale was expected to touch 6 million units by end of calendar 2010 in India. [25]

*1) 3G will accelerate the Sale of Mobile Phones in India*

In 2008, India entered the 3G arena when Government owned Bharat Sanchar Nigam Limited (BSNL) launched its 3G enabled mobile and data services. Later, Mahanagar Telephone Nigam Ltd (MTNL) also launched its services in Delhi and Mumbai. The private sector service providers such as Tata Docomo, Reliance Communications, Airtel, and Vodafone have also launched its 3G services.

3G enhances services such as multimedia and high speed mobile broadband. It equips the average mobile user with the ability to watch live TV on his/her mobile handset. One can also enjoy services such as live streaming, download of videos for educational or leisure purposes, news, current affairs and sport content and video messaging all in addition to the usual voice calling facility.

According to a recent forecast from the Wireless Intelligence, a service of trade group GSMA Ltd., India is all set to have 150 million 3G connections by the year 2014. The growing disposable income, reducing prices of all variety of mobile handsets, expanding penetration of 3G and reduced call and data rates has pushed the sale of mobile phones in India. It is expected to surpass China in near future and will stand first in terms of mobile subscribers.

## III. MOBILE PHONE AS HAZARD: COMPONENTS WITH CONSTITUENT & THEIR HEALTH EFFECTS

The Environmental Literacy Council gives a list of the components in cell phone. The list covers the basic components of most cell phones [10]. However some of the components may vary by individual cell phones. The traces of these components may reach inside the human body while the regular handling or by inhaling the dangerous fumes during the unsafe disposal of mobile waste (m-waste).

1. **Screen**, a user interface which is usually a liquid crystal display (LCD).
   - It contains lead, mercury, plastic etc.
2. **Green board**, the chips and electronic components which allow the cell phone to function properly.
   - It contains lead, nickel, zinc, beryllium, tantalum, coltan, copper, gold, and other metals.
3. **Battery**, a device which powers the phone.
   - It contains lead, acid, nickel, cobalt, zinc, cadmium, lithium and copper etc.
4. **Casing & Keypad,** the essential parts of a mobile phone.
   - It contains plastics including PVC & brominates flame retardants.
5. **Adapter**, a device to charge the cell phone's battery.
   - It contains plastics including PVC, ceramic capacitor, electrolytic capacitor etc. [19]

Table I. Worldwide Mobile Device Sales to End Users in 2010 (Thousands of Units) [23]

| Company | 2010 Units | 2010 Market Share (%) | 2009 Units | 2009 Market Share (%) | Difference in % from 2009 - 2010 |
|---|---|---|---|---|---|
| Nokia | 461,318.2 | 28.9 | 440,881.6 | 36.4 | 4.64 |
| Samsung | 281,065.8 | 17.6 | 235,772.0 | 19.5 | 19.21 |
| LG Electronics | 114,154.6 | 7.1 | 121,972.1 | 10.1 | -6.41 |
| Research In Motion | 47,451.6 | 3.0 | 34,346.6 | 2.8 | 38.16 |
| Apple | 46,598.3 | 2.9 | 24,889.7 | 2.1 | 87.22 |
| Sony Ericsson | 41,819.2 | 2.6 | 54,956.6 | 4.5 | -23.91 |
| Motorola | 38,553.7 | 2.4 | 58,475.2 | 4.8 | -34.07 |
| ZTE | 28,768.7 | 1.8 | 16,026.1 | 1.3 | 79.51 |
| HTC | 24,688.4 | 1.5 | 10,811.9 | 0.9 | 128.34 |
| Huawei | 23,814.7 | 1.5 | 13,490.6 | 1.1 | 76.53 |
| Others | 488,569.3 | 30.6 | 199,617.2 | 16.5 | 144.75 |
| **Total** | **1,596,802.4** | **100.0** | **1,211,239.6** | **100.0** | **31.83** |

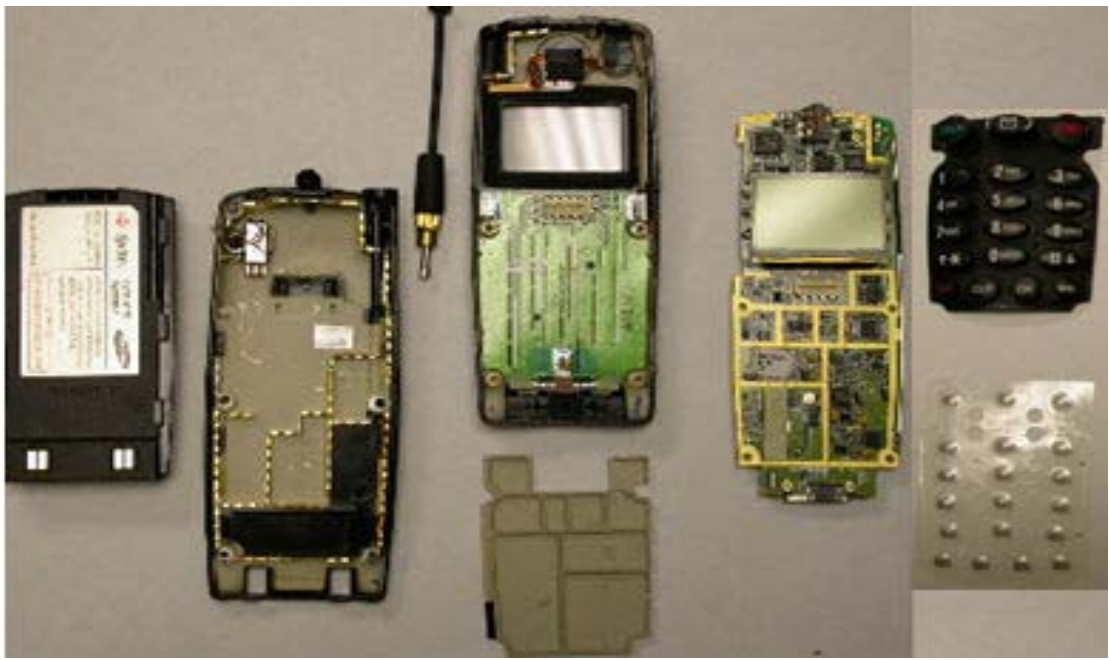

Fig.1 Typical components of a mobile phone. [26]

Table II.  Mobile Phone Component's Constituent and Their Health Effect [12][16][17][19].

| Constituent | Health effects |
|---|---|
| Lead (Pb) | • Damage to central and peripheral nervous systems, blood systems and kidney.<br>• Affects brain development of children. |
| Mercury (Hg) | • Chronic damage to the brain.<br>• Respiratory and skin disorders due to bioaccumulation in fishes. |
| Beryllium(Be) | • Develops carcinogenic (lung cancer) and skin diseases such as warts.<br>• Inhalation of fumes causes chronic beryllium disease or beryllicosis. |
| Plastics including PVC | • Burning produces dioxin. It causes reproductive and developmental problems.<br>• Interfere with regulatory hormones & damage to immune system. |
| Brominated flame retardants (BFR) | • Disrupts endocrine system functions. |
| Cadmium (Cd) | • Toxic irreversible effects on human health, accumulates in kidney and liver.<br>• Causes neural damage. |
| Lithium (Li) | • Shortness of breath, Cough, vomiting & weakness. |
| Cobalt (Co) | • Vomiting and nausea & Vision problems.<br>• Heart problems & Thyroid damage. |
| Nickel (Ni) | • Birth defects & Lung embolism.<br>• Allergic reactions such as skin rashes & Heart disorders. |
| Zinc (Zn) | • Too much zinc can cause stomach cramps, skin irritations, vomiting, nausea & anemia. |

## IV. DISASTROUS SCENARIO OF MOBILE PHONES WASTE

Discarded mobile phones create an avalanche of toxic e-waste. According to British newspaper 'The Independent', there are already 11,000 tons of unused cellular phones in the United Kingdom that have not yet been disposed of. These electronic products are made with highly toxic metals and other chemicals that leach into the earth when discarded. [5]

ABI Research (a market intelligence company) estimates that, in addition to shorter handset replacement cycles and a greater demand for cheaper phones will cause the recycled handset market to be worth $3 billion by 2012, with recycled phone shipments numbering above 100 million. [5]

Hard-rock mining of copper, silver, gold and other materials extracted from electronics is considered far more environmentally damaging than the recycling of those materials.

Guiyu in the Shantou region of China, Delhi and Bangalore in India as well as the Agbogbloshie site near Accra, Ghana have electronic waste processing areas. Uncontrolled burning, disassembly, and unsafe disposal cause a variety of environmental problems such as groundwater contamination, atmospheric pollution, or even water pollution either by immediate discharge or due to surface runoff (especially near coastal areas). It also creates health problems among those who are directly and indirectly involved in the methods of processing the m-waste.

Opponents of the trade argue that developing countries utilize methods that are more harmful and more-wasteful. An expedient and prevalent method is simply to toss equipment onto an open fire, in order to melt plastics and to burn away unvaluable metals. This releases carcinogens and neurotoxins into the air, contributing to acrid and lingering smog. These noxious fumes include dioxins and furans. Bonfire refuse can be disposed of quickly into drainage ditches or waterways feeding the ocean or local water supplies. [1]

The raw materials in a cell phone, such as gold, copper coils, aluminium and other metals are worth money but to extract these, the printed boards are basically cooked; releasing arsenic, mercury, lead and other toxins which harm the body. Inhaling, or regular handling of e-waste can result in damage to the brain, nervous system, lungs, and kidneys. Dr Venkatesh, in his address to a Hazardous Materials Seminar held recently in Bangalore explained that the estimated costs associated with lead poisoning amongst children in India are over $600 million per year. [2]

In China, India, Ghana and other developing regions thousands of men, women, and children are employed in highly polluting areas using primitive and unsafe recycling technologies to extract the metals, toners, and plastics from cell phones and other electronic waste. Recent studies show that 7 out of 10 children in these regions have too much lead in their blood. [1]

A recent study by the Chittaranjan National Cancer Institute, Kolkata, found that people in Delhi are about twice as likely to suffer from lung ailments as those in the countryside. Doctors blame on smelting the huge amount of electronic and mobile waste for the increasing lung ailments among the poor workers. [27]

## V. REGULATION AND POLICIES: OTHER COUNTRIES V/S INDIA

Recycling and disposal of e-waste may involve significant risk to workers and communities. A great care must be taken to avoid unsafe exposure in recycling operations and leaching of material such as heavy metals from landfills and incinerator ashes. Scrap industry and USA EPA (Environment Protection Agency) officials agree that materials should be managed with caution, and environmental dangers of unused electronics have not been exaggerated. [1]

### A. Regulation Regarding E-Waste in Developed World

The U.S. government supports many local and statewide cell phone recycling programs. These programs reuse parts of cell phones that are useful in other electronic devices. They also refurbish cell phones and other electronic devices for use in schools or for low-income families who cannot afford new phones. Government-supported programs (like donation, E-recycling etc. [20]) may be found on the U.S. Environmental Protection Agency website (see Resources). [9]

According to Thomsen, about 100 million people upgrade to new phones each year in Europe alone, even though the average handset has a life of 5 years.

The European Union (EU), Japan, South Korea, Taiwan and several states of the USA have introduced legislation making producers responsible for their end-of-life products. The EU has banned the use of certain hazardous substances in electrical and electronic products from July2006, to facilitate safer recycling. [13]

California has taken recycling one step further than the EPA (Environmental Protection Agency) with the introduction of the California Cell Phone Recycling Act in 2004. The act requires cell phone retailers to accept all cell phones from consumers for recycling. As a result, about 3.6 million phones or 25% of the phones sold in California were reused in 2008. [10]

Other U.S. states considering similar legislation include Illinois, Mississippi, New Jersey, New York, Vermont and Virginia, while the Canadian provinces of British Columbia, Alberta, Saskatchewan and New Brunswick are likely to jump on the mandatory cell phone recycling bandwagon soon. [11]

### B. Regulation Regarding E-Waste in India

The unfortunate part is that while developed countries have a proper system for recycling of disposed e-devices, such a system is lacking in India. It's not just about a system, even awareness on recycling e-waste is lacking in the second largest mobile market in the world. In India, there are no specific environmental laws or guidelines for m-waste or e-waste. None of the existing environmental laws have any direct reference to electronic waste or refer to the way it is handled as being hazardous. However, as some components of electronic waste fall under the 'hazardous and 'non-hazardous' ( Hazardous waste that poses substantial or potential threats to public health or the environment [15] such as batteries, switches etc, and non-hazardous waste like plastic, circuit board etc. [14]) waste categories; they are covered under the purview of 'The Hazardous Waste Management Rules, 2003'. This regulation defines hazardous waste as "any waste which by reason of any of its physical, chemical, reactive, toxic, flammable, explosive or corrosive characteristics causes danger or is likely to cause danger to health or environment, whether alone or when on contact with other wastes or substances." As per the guidelines for environmentally sound management of e-waste, Maharashtra ranks first followed by Tamil Nadu, Andhra Pradesh, Uttar Pradesh, West Bengal, Delhi, Karnataka, Gujarat, Madhya Pradesh and Punjab in the list of e-waste generating states in India. In these guidelines, the Ministry of Environments and Forests' central Pollution Control Board has proposed the extended producer responsibility (EPR) as an environment protection strategy. This makes the producer responsible for the entire life cycle of the product, especially for take back, recycle and final disposal. Thus, the producers' responsibility is extended to the post consumer stage of the product life cycle. This needs to be included in the legislative framework making EPR a mandatory activity associated with the production of electronic and electrical equipment over a period of time. [6]

## VI. INDUSTRIES' INITIATIVE IN INDIA

Many organizations in India are trying to put in order the way recycling is done in the country. These organizations collect e-waste through their collection centres and transport them to recycling plants. Once the scrap reaches a plant, metallic and non metallic components are separated. Telecom giants such as Nokia, LG and Tata Teleservices (TTSL) have now started generating awareness about the issue. Nokia has started a 'take back' scheme in various cities wherein mobile phone users can dispose their used handsets and accessories, regardless of the brand, at recycling bins in Nokia Priority Dealers and Nokia Care Centers. Nokia had collected close to 16 tones of e-waste (mobile phones and accessories) till April this year as part of their 'take back' campaign started in January 2009. Nokia's 'Planet Ke Rakhwale' community already has a member base of 20,000 people. On the other hand, recycling companies such as Attero Recycling work with various companies including telecom giants such as LG and TTSL. They also try to touch base with the informal sector and try to minimize damage to the environment and to human health by open air burning [6]. Dell, E-Parisaraa, Green India Recycling Pvt. Ltd., Trishyiraya Recycling India Private Limited and many other companies also contributing for collecting e-waste.

## VI. CONCLUSION & SUGGESTIONS

The problem of accumulating m-waste must be addressed immediately otherwise it will lead to deceases and casualties of our people. Not only the Government and Industries, but the Citizens also have a very important role to play. We are suggesting some immediate steps to tackle the proper disposal of m-waste. [6].

### A. Responsibilities of the Government

1. Governments should set up regulatory agencies in each district, which are vested with the responsibility of coordinating and consolidating the regulatory functions of the various government authorities regarding hazardous substances.
2. Existing laws concerning e-waste disposal be reviewed and revamped. A comprehensive law that provides mobile and e-waste regulation and management and proper disposal of hazardous wastes is required. Such a law should empower the agency to control, supervise and regulate the relevant activities of government departments.
3. Control risks from manufacture, processing, distribution, use and disposal of electronic wastes.
4. Encourage beneficial reuse of e-waste. Also set up programs to promote recycling among citizens and businesses.
5. Government should enforce strict regulations against dumping e-waste in the country by outsiders and industries which do not practice waste prevention and recovery in the production facilities. Where the laws are flouted, stringent penalties must be imposed.
6. Government should encourage and support NGOs and other organizations to involve actively in solving the nation's e-waste problems.
7. Government should explore opportunities to collaborate with manufacturers and retailers to provide recycling services [12].
8. Innovative programs should be encouraged, like sending the SMS regarding the safe disposal to all those mobile users who are using the same handset for a longer period of time.

### B. Responsibility and Role of Industries

1. Companies can and should adopt waste minimization techniques, which will make a significant reduction in the quantity of e-waste generated and thereby lessening the impact on the environment. It is a "reverse production" system that designs infrastructure to recover and reuse every material contained within e-wastes metals such as lead, copper, aluminum, gold, plastics, glass and wire. Such a "closed loop" manufacturing and recovery system offers a win-win situation for everyone. Less of the Earth will be mined for raw materials, and groundwater will be protected.
2. Manufacturers, distributors, and retailers should undertake the responsibility of recycling/disposal of their own products.
3. Standardize components for easy disassembly.
4. Utilize technology sharing particularly for manufacturing and de-manufacturing.
5. Encourage / promote / require green procurement for corporate buyers.
6. Use label materials to assist in recycling (particularly plastics) [12].

### C. Responsibilities of a Citizen

1. E-wastes should never be disposed with garbage and other household waste. This should be segregated at the site and sold or donated to various organizations.
2. Customers should opt for upgrading their cell phone to the latest versions rather than buying new equipments.
3. NGOs should adopt a participatory approach in management of e-waste [12].

### D. Safe Technique for Metal Recovery & Encourage the Reuse

1. Waste can be recovered on-site, or at an off-site recovery facility, or through inter industry exchange. To reclaim the waste material, a number of physical and chemical techniques are available such as reverse osmosis, electrolysis, condensation, electrolytic recovery, filtration, centrifugation etc. For example, a printed-circuit board manufacturer can use electrolytic recovery to reclaim metals from copper and tin-lead plating bath [12].
2. The existing and potential technologies that might be used for the metal recycling include mechanical processing, pyrometallurgy, hydrometallurgy, biohydro-metallurgy or a combination of these techniques. Of these techniques, hydrometallurgical approach is often used due to energy efficient and flexible to a variation in the metal contents [3].

To encourage phone reuse, Green Mobile (Green Mobile in partnership with the Friends of the Earth -which is an organization dedicated to the care of our planet; and Environmental Investigation Agency (EIA) has created the UK's first environmentally friendly mobile phone service. By joining Green Mobile, a mobile user agree to hold on his/her existing handset for just one more year and in return organization donate £15 to EIA or Friends of the Earth [28]) asks new customers to keep using their old handset and rewards them with a lower rate than can be offered by companies that subsidize new phones each year [5].

The prevalence of recycled phones is expected to increase as the problem of e-waste enters the public consciousness and stricter regulations force more companies to tackle the problem [5].